\documentclass[12pt,preprint]{aastex}

\usepackage{amsmath}
\usepackage{graphicx}
\usepackage{multirow}
\usepackage{color}
\usepackage{natbib}

\begin{document}

\title{Do Centaurs Preserve their Source Inclinations?}

\author{Kathryn Volk and Renu Malhotra}
\affil{Lunar and Planetary Laboratory, University of Arizona, Tucson, AZ 85721, USA.}
\email{kvolk@lpl.arizona.edu}

\begin{abstract}

The Centaurs are a population of small, planet-crossing objects in the outer solar system.  They are dynamically short-lived and represent the transition population between the Kuiper belt and the Jupiter family short-period comets.  Dynamical models and observations of the physical properties of the Centaurs indicate that they may have multiple source populations in the trans-Neptunian region.  It has been suggested that the inclination distribution of the Centaurs may be useful in distinguishing amongst these source regions.  The Centaurs, however, undergo many close encounters with the giant planets during their orbital evolution; here we show that these encounters can substantially determine the inclination distribution of the Centaurs.  Almost any plausible initial inclination distribution of a Kuiper belt source results in Centaurs having inclinations  peaked near $10-20^{\circ}$.  Our studies also find that the Kuiper belt is an extremely unlikely source of the retrograde Centaur that has been observed. 
\end{abstract}

\section{Introduction}\label{s:Introduction}

The group of minor planets known as Centaurs represent the dynamical link between the reservoirs of icy objects in the outer solar system and the short-period Jupiter family comets in the inner solar system.  Because the orbits of the Centaurs cross those of the outer planets, Centaurs are dynamically short-lived. \citet{Tiscareno2003} numerically integrated the orbits of 53 observed Centaurs and found that their median dynamical lifetime was only 9 Myr.  This means that there must exist a long-lived source for this transient population.  Attempts to identify the Centaur sources have largely focused on dynamical models of the potential source regions.  Early studies to identify a source region for the short period comets suggested that the long period comets (originating from the nearly isotropic distant Oort Cloud) could be captured into short period orbits by means of planetary perturbations, and that the higher capture probabilities at low inclinations could explain the prograde, low inclinations of the short period comets \citep{Everhart1972}.  But \citet{Fernandez1980} and \citet{Duncan1988} showed that Everhart's mechanism is a very inefficient way to produce Centaurs and short period comets from Oort Cloud comets, suggesting that a more proximate, lower-inclination population just beyond Neptune would be a more likely source region.  With the subsequent discovery of the Kuiper Belt and its complex dynamical structure, recent studies have focused on the dynamical subclasses of the Kuiper belt  as the sources that resupply the Centaurs and the Jupiter family comets: \cite{Levison1997}, \cite{Volk2008} and \cite{DiSisto2007} modeled the scattered disk source; \cite{DiSisto2010} and \cite{Morbidelli1997} modeled the Plutinos (objects in the 3:2 mean motion resonance with Neptune) as a source of the short period comets; the Trojan populations of Neptune and/or Jupiter (objects in the 1:1 mean motion resonance with Neptune or Jupiter) were studied by \cite{Horner2010} and \cite{Horner2006}.  The Oort cloud has also been revisited as a possible source \citep{Emelyanenko2005,Brasser2012}. These models have identified dynamical pathways as well as the flux per unit source population from each source, but it remains unclear which, if any, source population dominates the flux of new Centaurs.  We do not yet have strong enough observational estimates of the number of small, comet-sized (1--10 km) bodies in each of the proposed source regions to be able to estimate the absolute number of new Centaurs and/or Jupiter family comets from each region \citep{Bernstein2004,Levison1997s,Volk2008,Fraser2010}.  

Another approach is to examine the dynamical and/or physical properties of the observed Centaurs and compare them to the same properties in the source region.  Because the Centaurs have orbits that are much more favorable for observations than the orbits of trans-Neptunian objects (TNOs), they have been the subject of several spectroscopic and photometric studies (\citet{Tegler2008} and references therein, \citet{Barucci2011}).  Interestingly, the Centaurs have a bimodal color distribution \citep{Tegler2008}, which was not found in the early studies of (brighter, larger) TNOs.  New studies indicate that the color distribution of TNOs is size dependent and that a bimodal color distribution occurs in the population of smaller, dynamically excited TNOs \citep{Fraser2012,Peixinho2012}.  These observations give us some clues about the origins of the Centaurs, but there are not yet large enough sample sizes to draw any firm conclusions.  Furthermore, due to the inherent faintness of TNOs as observed from Earth, it is unclear that we will be able to observe sufficient numbers of smaller, Centaur-sized TNOs in the near future to improve this state.

\begin{figure}
   \centering
   \includegraphics{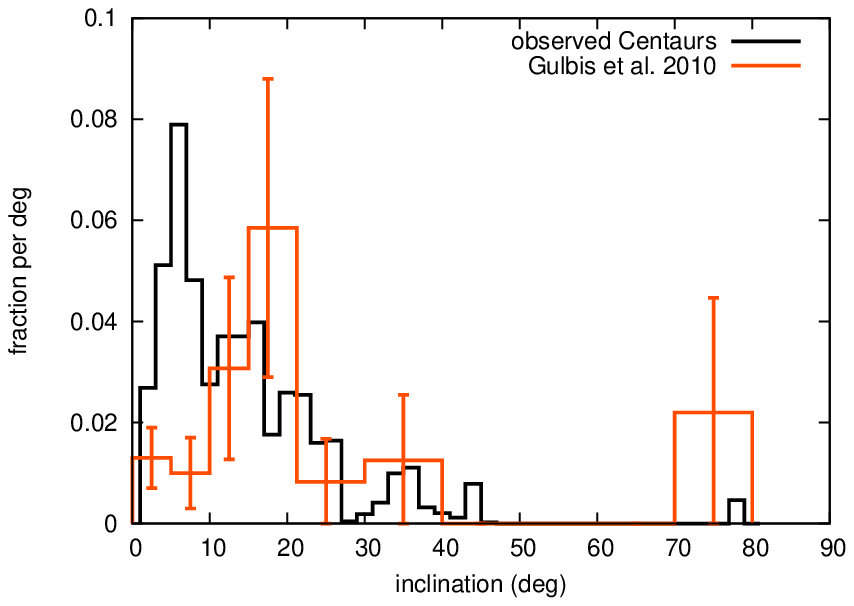} 
   \caption{The inclination distribution of the observed Centaurs (data from the Minor Planet Center website) and the debiased inclination distribution for the Centaurs from the Deep Ecliptic Survey \citep{Gulbis2010}.}
   \label{f:known_centaurs}
\end{figure}

The dynamical properties of the Centaurs might be a possible way of identifying the sources.  The dynamics of the observed Centaurs have been modeled by \citet{Tiscareno2003} and \citet{Bailey2009}; their orbital evolution is dominated by chaotic diffusion and the frequent close encounters with the outer planets, which cause their orbital elements to evolve rapidly.  It has been suggested (e.g. \citet{Gulbis2010}) that, despite their strongly chaotic evolution, the orbital inclinations of the Centaurs might preserve memory of their source regions.  There are currently sufficient observations of TNOs to identify several dynamical subclasses and to calculate inclination distributions separately for the classical Kuiper belt, the scattered disk, and the resonant populations \citep{Brown2001,Elliot2005,Gulbis2010}.  The inclination distribution of the Centaurs is less well defined due to the relatively small number, 108, of observed Centaurs\footnote{www.minorplanetcenter.net/iau/lists/t\_centaurs.html}.  The observed inclination distribution of these objects is shown in Fig.~\ref{f:known_centaurs} along with the debiased inclination distribution from the Deep Ecliptic Survey \citep{Gulbis2010}. If the Centaurs' inclination distribution does retain some memory of the source region's inclination distribution, that could be an important means for linking the Centaurs to a particular dynamical subclass of the Kuiper belt or to the Oort cloud.

In this work, we investigate the extent to which inclinations of Centaurs may preserve the inclination distribution of their sources.  We present the results of numerical simulations of hypothetical Centaurs generated from the Kuiper belt subclasses.  We follow the orbital evolution of these hypothetical Centaurs until they either transition onto inner solar system orbits or they are ejected from the giant planet region.  During this evolution, our simulated Centaurs experience many close encounters with the four outer planets, each of which induces some change in the orbital elements of the Centaurs.  We use these simulations to calculate the average outcomes of planetary encounters, including the average change in a Centaur's inclination as a function of close encounter distance.  Using these average encounter outcomes, along with statistics about close encounter frequency as a function of encounter distance and the dynamical lifetimes of Centaurs, we compute the dynamically evolved inclination distribution expected of Centaurs as a function of the initial inclination distribution of the source population.  We then discuss the implications of these results for identifying the dominant Centaur source(s). 

\section{Numerical Simulations}\label{s:sims}

To explore the evolution of inclinations in the Centaur region, we performed a numerical integration of test particles entering the planet-crossing region from a trans-Neptunian source region.  We chose source region parameters based on the classical Kuiper belt (CKB) and the scattered disk (SD) because these dynamical classes are widely regarded as the most likely sources of the Centaurs \citep{Levison1997,DiSisto2007,Volk2008,Lowry2008}.  This numerical simulation is not meant to be exactly representative of either the CKB or the SD, but is rather an exploration of the parameter space of these populations.  To that end, our test particles uniformly cover the following ranges in semimajor axis ($a$), perihelion distance ($q$), mean anomaly ($M$), argument of perihelion ($\omega$), and longitude of ascending node ($\Omega$):  
\begin{itemize}
\item $40$ AU $\le a \le$ $80$ AU
\item $30$ AU $\le q \le$ $32$ AU
\item $0$ $\le$ $M,\Omega,\omega$ $<$ $2\pi$.
\end{itemize}
To explore a large range in inclination, the test particles were divided into 24 inclination bins (200 test particles per bin) with the following bin widths:
\begin{itemize}
\item $1^{\circ}$ for $0^{\circ}$ $\le i <$ $10^{\circ}$
\item $2^{\circ}$ for $10^{\circ}$ $\le i <$ $30^{\circ}$
\item$5^{\circ}$ for $30^{\circ}$ $\le i <$ $50^{\circ}$
\end{itemize}
and inclinations were assigned uniformly within each bin.  The choice to limit perihelion distances to the $30-32$ AU range was made to manage computational time.  Test particles with these $q$ values will either enter the Centaur population (which we define $5 < q < 30$ AU) or be ejected from the solar system early in the simulation, minimizing the necessary total length of the simulation.  In previous scattered disk simulations \citep{Volk2008}, we found that the inclination distribution is preserved as test particles evolve from $q>33$ AU to $q\sim30$ where they start to have close encounters with Neptune; the test particles' inclinations on our initially nearly Neptune-crossing orbits are a good representation of their larger $q$ source region inclinations.  We limit the initial semimajor axes to $<80$ AU to conserve computational time because test particles starting at larger values of $a$ will take longer to evolve onto Centaur-like orbits.  We show later in this section that the dynamical properties we are interested in do not depend on this limit in $a$.

The integration was performed using the swift\_rmvs3 code in the SWIFT software package\footnote{http://www.boulder.swri.edu/$\sim$hal/swift.html \citep{Levison1994}}, which is capable of integrating test particles through arbitrarily close encounters with the massive planets.  For each simulation we include the four outer planets and the sun as massive bodies (the sun's mass is augmented with the mass of the terrestrial planets), and the hypothetical Centaurs are included as massless test particles.   We use astep size of one year, and we remove test particles if they achieve a heliocentric distance inside Jupiter's orbit or beyond 1000 AU, or if they impact one of the giant planets.  The inner boundary of the simulation was chosen to conserve computational time; following test particles into the inner solar system would require the inclusion of the terrestrial planets and the use of a much shorter  integration step size.  This choice means that we are not including in our fictitious Centaur population those objects that enter the inner solar system, become Jupiter family comets, then evolve back onto Centaur orbits.  Previous work has shown that once test particles evolve to Jupiter crossing orbits they are either ejected from the solar system or they become short-period comets which are then ejected from the solar system on very short timescales ($\sim 10^5$ yrs) \citep{Tiscareno2003,DiSisto2007,Bailey2009}; this means that for time-weighted distributions in the Centaur region, our inner boundary will have almost no effect on the results of our simulations.  The simulation was run for 300 Myr because this is many times longer than the expected dynamical lifetime ($\sim$ 10 Myr) for most Centaurs and longer than even the very long lived Centaurs simulated by \citet{Tiscareno2003}, \citet{DiSisto2007}, and~\citet{Bailey2009}.  In the analysis of our simulations, we are particularly interested in the dynamical lifetimes of the Centaurs produced from our source population, as well as the close encounters they have with the four outer planets and the evolution of the inclinations of the test particles.  In Section~\ref{s:map} we discuss how we use the results of this simulation to map arbitrary source inclination distributions to their Centaur inclination distribution without the need for additional numerical simulations.  

\subsection{Planetary encounter statistics}\label{ss:ce}

\begin{figure}
   \centering
   \includegraphics{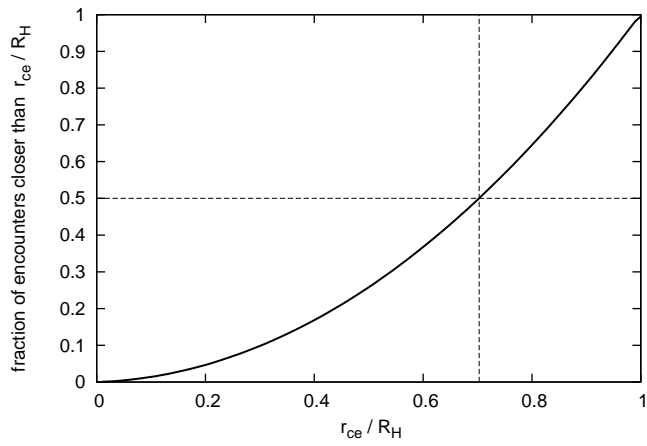} 
   \caption{Cumulative fraction of encounters (solid line) occurring at closest approach distances less than $r_{ce}/R_H$.  The dashed lines indicate the approach distance outside of which 50\% of encounters occur.}
   \label{f:rce}
\end{figure}

The output of our simulation includes the orbital histories of all the test particles as well as the detailed outcomes of more than 1.8 million planetary encounters  with Neptune, Uranus, Saturn and Jupiter.  A planetary encounter is defined as approaching within one Hill radius, $R_H$, of a planet, 
\begin{equation}\label{eq:rhill}
R_H = a_p \left(\frac{M_p}{3 M_{\sun}}\right)^{1/3},
\end{equation}
where $a_p$ is the planet's semimajor axis, $M_p$ is its mass, and $M_{\sun}$ is the mass of the sun.  
During its lifetime as a Centaur, each test particle has an average of 420 planetary encounters before it either enters the inner solar system or is ejected from the solar system to distances greater than 1000 AU.  Fig.~\ref{f:rce} shows the distribution of closest approach distances, $r_{ce}$, scaled to the encountered planet's Hill radius.
We find that 50\% of the encounters in the simulation occur at separations larger than $0.7 R_H$.  The distribution of closest approach distances follows a power law $\sim(R_H/r_{ce})^2$. This is consistent with expectations from geometrical arguments that neglect gravitational focusing due to the planet.

\begin{figure}
   \centering
   \includegraphics[width=6.75in]{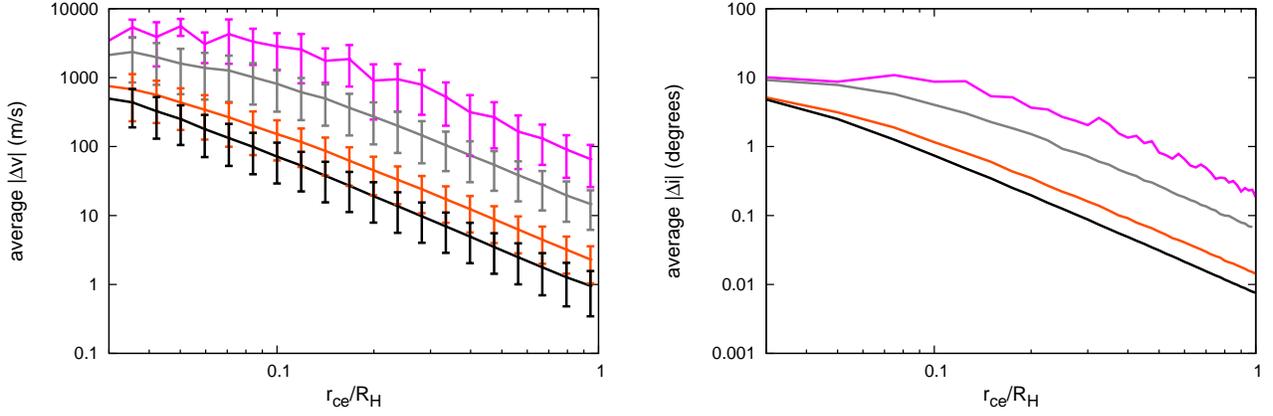} 
   \caption{Average $|\Delta v|$ per encounter (left panel) and average $|\Delta i|$ per encounter (right panel) vs the scaled closest approach distance for the simulation results.  Results for encounters with Jupiter are shown in purple (largest $|\Delta v|,|i|$), Saturn in grey, Uranus in orange, and Neptune in black.   The error bars in the left panel are the rms variations in $|\Delta v|$.}
   \label{f:dv_di}
\end{figure}

We also examine the change in a test particle's heliocentric velocity, $\Delta v = |\vec{v_f} - \vec{v_i}|$, and the change in inclination, $|\Delta i|$, imparted by each planetary encounter. These are plotted as a function of $r_{ce}/R_H$ for each planet in Fig.~\ref{f:dv_di}. The error bars indicate the rms variations for each bin in $r_{ce}/R_H$.  The majority of encounters have only very small effects on the test particles' orbits because they occur at relatively large separations.  
For $r_{ce}/R_H\gtrsim0.1$, the average changes, $\langle|\Delta v|\rangle$ and $\langle|\Delta i|\rangle$, as a function of encounter distance can be fit to a power law in the scaled closest approach distance, the mass of the planet, and the semimajor axis of the planet:
\begin{equation}\label{eq:dv}
\left<|\Delta v|\right> = 6.24\times10^3 (ms^{-1}) \left(\frac{M_p}{M_\sun}\right)^{0.18} \left(\frac{a_p}{AU}\right)^{-2.1} \left(\frac{r_{ce}}{R_H}\right)^{-2},
\end{equation}
and
\begin{equation}\label{eq:di}
\left<|\Delta i|\right> = 9.7^{\circ} \left(\frac{M_p}{M_\sun}\right)^{0.17} \left(\frac{a_p}{AU}\right)^{-1.6} \left(\frac{r_{ce}}{R_H}\right)^{-2}.
\end{equation}
The best-fit parameters in the above equations are based on the simulated close encounters with Saturn, Uranus, and Neptune; Jupiter encounters were excluded from the fits because the inner boundary of the simulation does not allow us to follow a sufficient number of test particles through Jupiter encounters.  We also note that the range of planet mass is quite small: we have only three planets in our fit, Saturn, Uranus and Neptune, and the latter two have very similar masses, which means that the planet mass dependence of our best-fit, Eq.~\ref{eq:dv}, is not very well determined. 

It is interesting to compare these scalings with analytic estimates of $\Delta v$.  In the limit of nearly circular, coplanar orbits very close to a planet, let the particle's semimajor axis be $a=a_p(1+x)$.  Then, the closest approach distance is given by $r_{ce} = xa_p$, and the relative angular speed is $n-n_p=n_p(1+x)^{-2/3}-n_p\approx -{2\over3}xn_p$, where $n$ and $n_p$ are the mean motions of the test particle and planet respectively.  Let us define the encounter time, $t_{enc}$, to be the time it takes to move an azimuthal angular distance $2x$ relative to the planet; then the encounter time is given by
\begin{equation}
t_{enc} = \left|\frac{2x}{n-n_p} \right | \approx{4\over3n_p}.
\end{equation}
The velocity kick from a planetary encounter can then be estimated as
\begin{equation}
|\Delta v| \approx \left |-\frac{G M_p}{(xa_p)^2} \right | t_{enc} = \frac{4}{3}\frac{M_p}{M_\sun}(GM_\sun)^{1/2}a_p^{3/2}r_{ce}^{-2}.
\end{equation}
Scaling the encounter distance to the planet's Hill radius, we find that nearly circular test particles would experience a velocity kick that scales with planet mass and semimajor axis in the following way
\begin{equation}\label{eq:cdv}
\Delta v \propto \left( \frac{M_p}{M_\sun} \right)^{1/3} \left(\frac{a_p}{AU}\right)^{-1/2}\left(\frac{r_{ce}}{R_H}\right)^{-2}.
\end{equation}
The orbits of the Centaurs are neither circular nor coplanar, but this serves as a useful comparison for the simulation results.  The simulation results are consistent with the analytic estimate (Eq.~\ref{eq:cdv}) for the scaling of $\Delta v$ with $r_{ce}$, but not for the scaling in planet mass and semimajor axis.  These differences are partially due to the substantial eccentricities and inclinations of the Centaurs in our simulation; this affects their encounter velocities and encounter times.   The planet mass and semimajor axis scalings could also be dependent on the architecture of the giant planets in our solar system; a different planet configuration may yield a different scaling.

\subsection{Dynamical lifetimes and inclination changes}\label{ss:lt_di}

\begin{figure}
   \centering
   \includegraphics[width=6.75in]{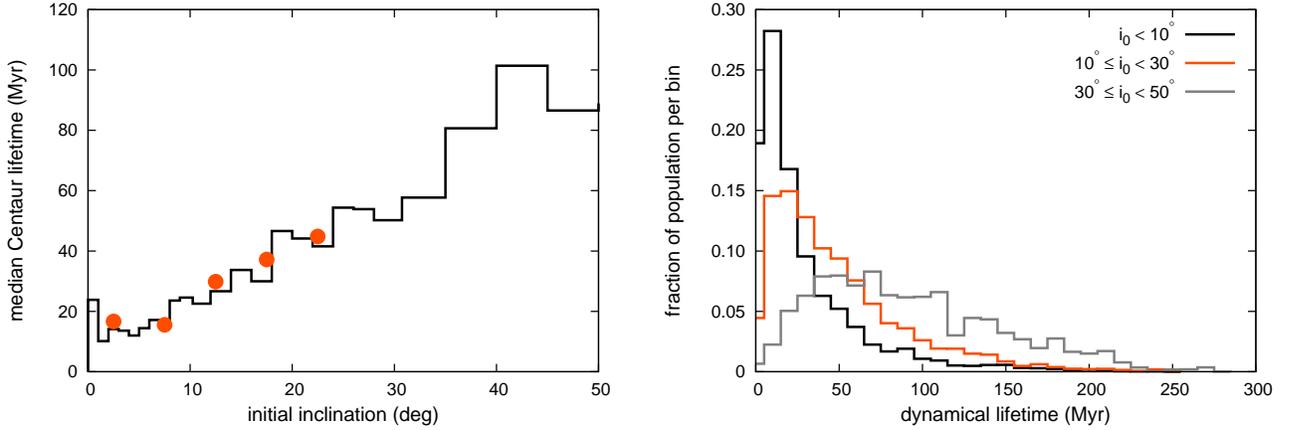} 
   \caption{ Left panel: median dynamical lifetime as a Centaur vs. initial inclination. The bin sizes vary such that there are roughly equal numbers of test particles contributing to each bin. The solid line shows the simulation results described in Section~\ref{ss:lt_di} and the orange dots show the lifetimes of Centaurs originating in the 3:2 resonance with Neptune (see Section~\ref{s:discussion}).  Right panel: distribution of Centaur lifetimes for three different initial inclination ranges.}
   \label{f:lt}
\end{figure}

We also examine the dynamical lifetimes of the Centaurs in our simulation.  Here we define the dynamical lifetime as the total time spent in the perihelion distance range $5 < q < 30$ AU during the simulation\footnote{Some test particles evolve out of the defined Centaur perihelion range and then back in.  However these excursions are generally short compared to the total dynamical lifetime, so their inclusion or exclusion from the computation of the dynamical lifetime does not change the results reported here.}.  Fig.~\ref{f:lt} shows the median dynamical lifetimes, $\tau$, of our simulated Centaurs as a function of their initial inclination in the source region; the binning scheme is the same as that for the initial conditions.  Only test particles that entered the Centaur region ($5 < q < 30$ AU), and were subsequently removed from the simulation (at the inner or outer boundaries, or after a collision with a planet) are included in the calculation of the dynamical lifetimes.  Even with our initial, near-Neptune perihelion distances, some test particles don't enter the Centaur population until late in the simulation; because these test particles are still evolving as Centaurs when the simulation was stopped, we do not have accurate values of $\tau$ for them, so they are excluded.

The time spent as a Centaur increases with initial inclination, with the highest inclination test particles spending about 4 to 5 times longer as Centaurs than the lowest inclination test particles.  This is consistent with previous work by \citet{DiSisto2007} (see their Fig. 5), although the values of $\tau$ are slightly different because they use the mean value of $\tau$ while we use the median.  We prefer to use the median value for the dynamical lifetimes because the distribution of $\tau$ for any bin in initial inclination has a very long tail, which tends to skew the mean toward larger $\tau$.  Fig.~\ref{f:lt} also shows the distribution of $\tau$ for three different ranges in initial inclination.  The increase in $\tau$ with increasing initial $i$ is simply due to a decrease in the rate of planetary encounters because high-$i$ test particles spend little time near the plane of the planets.  The number of planetary encounters required to either transfer a test particle onto an orbit that enters the inner solar system or eject it does not depend on initial inclination (for the range of inclinations we tested), so the lower frequency of encounters corresponds to longer dynamical lifetimes.

\begin{figure}
   \centering
   \includegraphics{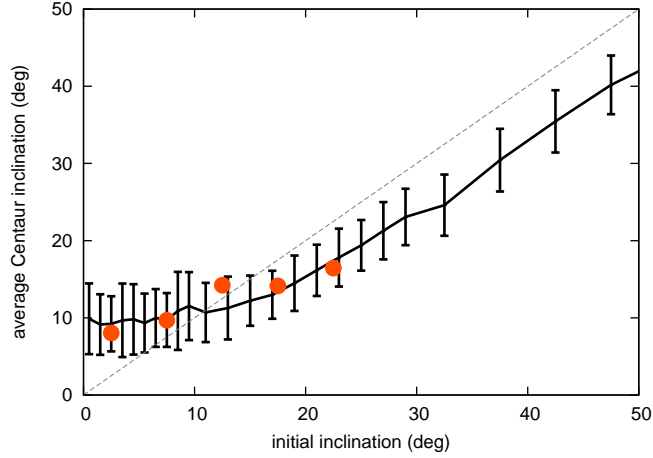} 
   \caption{Average inclination as a Centaur vs initial inclination in the source region (black line).  The error bars are the rms variation in i within each initial inclination bin, and the dashed line shows a one to one correlation for reference.  The orange dots show the average inclination vs initial inclination for Centaurs originating in the 3:2 resonance with Neptune (see Section~\ref{s:discussion}). }
   \label{f:icent}
\end{figure}

The net result of all these planetary encounters can be seen in Fig.~\ref{f:icent} which shows the average inclinations of the Centaurs in our simulation as a function of their initial inclinations.  Test particles that start with inclinations below $\sim 10^{\circ}$ have their inclinations nearly uniformly raised to $\sim 10^{\circ}$.  At higher initial inclinations, the planetary encounters have a net effect of slightly lowering the average inclinations.  

To see how dependent these results are on the semimajor axis distribution of our source region, we performed an additional simulation with 800 test particles spread uniformly in semimajor axis from $100-500$ AU with the same inclination and perihelion distance distribution as outlined above.  We calculated the average inclinations and median dynamical lifetimes for the resulting Centaurs and performed a Spearman rank correlation test to see how these two parameters depend on the initial semimajor axis.  This test yields a number in the range $[-1,1]$, with values near 0 indicating that the two variables being compared are not correlated (changes in one variable do not predictably result in changes to the other).  Comparing initial semimajor axis with Centaur lifetimes and with the average change in inclination during that lifetime, we find Spearman rank correlation coefficients of $\sim 10^{-2}$; this is consistent with no correlation between initial $a$ and either dynamical lifetime or $\Delta i$. This result may appear surprising, but can be understood as follows.  The evolution of Centaurs is nearly insensitive to their initial semimajor axis (for $a\gtrsim100$ AU) because such test particles on just-barely Neptune-crossing orbits will exhibit a random walk behavior in $a$ as they undergo many distant encounters with Neptune, while maintaining $q > 30$ AU; in general, we have found that test particles do not enter the Centaur region ($5 < q < 30$ AU) until they have random walked to semimajor axes less than $\sim100$ AU and that the inclinations of these test particles are maintained during this random walk. This means that the averaged results from our initial 2400 test particle simulation should be applicable to trans-Neptunian source regions with semimajor axis distributions that differ from our uniform $40\le a \le80$ AU test simulation.

\section{Mapping the Source Inclination Distribution to the Centaur Region}\label{s:map}

\begin{figure}
   \centering
   \includegraphics[width=6.75in]{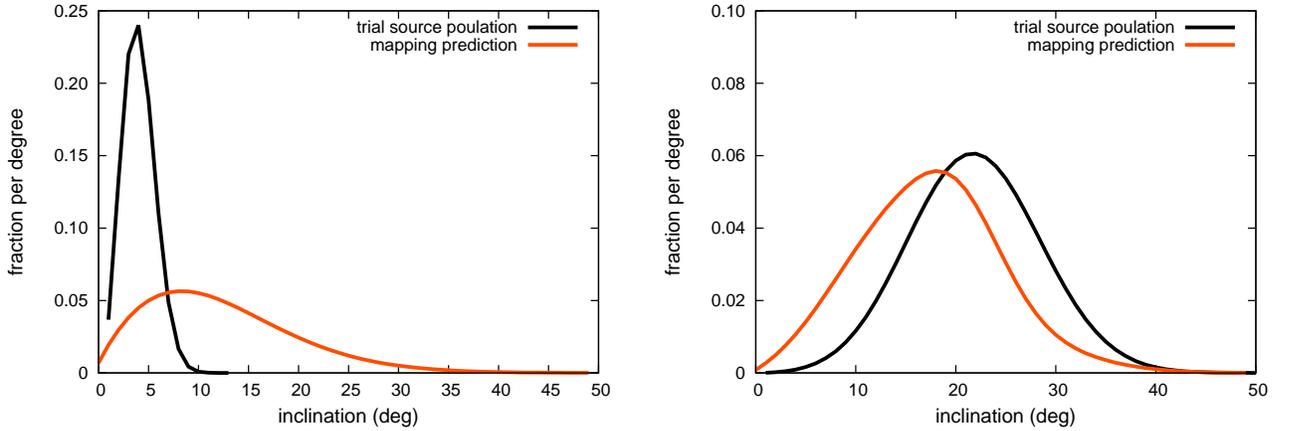} 
   \caption{Inclination distributions for a trial source population with an inclination distribution given by Eq.~\ref{eq:g} with $\mu=2^{\circ}$ and $\sigma = 2^{\circ}$ (left panel) and $\mu=19^{\circ}$ and $\sigma = 7^{\circ}$ (right panel) compared to the Centaur distributions given by the mapping procedure (Eq.~\ref{eq:map}). }
   \label{f:low_hi}
\end{figure}

The results of the simulation in Section~\ref{s:sims} tell us, on average, how the inclination and dynamical lifetime of a Centaur is related to its initial inclination.  As discussed in Section~\ref{s:sims}, these two key parameters are not much affected by the initial semimajor axis distribution of the test particles, at least within the range of semimajor axes typical for the Kuiper belt and the scattered disk.  This means that without worrying about the distributions of the other orbital elements, we can take an arbitrary source region inclination distribution and calculate what the $i$-distribution would be for the resulting Centaurs.  We divide the range of initial inclinations into bins (using the same scheme as in Section~\ref{s:sims}), and we model the inclinations of the Centaurs in each bin as a Gaussian multiplied by $\sin(i)$:
\begin{equation}\label{eq:g}
g(i,\mu,\sigma) = C \sin(i) \exp\left(\frac{-(i-\mu)^2}{2 \sigma^2}\right) ,
\end{equation} 
where $\mu$ and $\sigma$ are fit parameters determined from the simulation results and $C$ is a normalization constant; the values of these parameters are listed in Table~\ref{t:fits}.  Using these fits to the simulation results and the median dynamical lifetimes for each inclination bin, $\tau_n$, as a weighting factor (also listed in Table~\ref{t:fits}), we calculate the time-averaged Centaur inclination distribution as follows:
\begin{equation}\label{eq:map}
f_{centaur}(i) =  C \sum_{n=1}^{n_{bins}} \left[ g_n(i,\mu, \sigma) \tau_n  \frac{1}{(i_{n+1} - i_{n})}\int_{i_n}^{i_{n+1}} f_{source}(i) di\right],
\end{equation}
where  $i_n$ is the lower bound on the initial inclinations in bin $n$, and $C$ is a normalization factor, and $f_{source}$ is the source region inclination distribution.  Fig.~\ref{f:low_hi} shows the results of this mapping procedure for two trial source region inclination distributions; both source distributions take the form of a Gaussian multiplied by $\sin(i)$ with the left panel representing a very low-$i$ source region and the right panel representing a higher-$i$ source region.  For a very low-$i$ source region, the resulting Centaur $i$-distribution has a much larger spread and a larger mean $i$ than the source; the planetary encounters systematically raise the inclinations of the Centaurs.  Centaurs from a high-$i$ source region will have an inclination distribution similar to the source region, but skewed slightly toward lower $i$.

\section{Comparing the Mapping Procedure to the Results of Numerical Simulations}\label{s:compare}

To test the accuracy of the mapping procedure, we performed two additional numerical simulations:  one of a CKB-like source region and one of a SD-like source region.  The CKB-like source consisted of 2400 test particles with semimajor axes spread evenly over the range $40 AU < a < 50 AU$ with an inclination distribution given by \citet{Brown2001}'s debiased best fit to the observed classical belt objects: 
\begin{equation}\label{eq:brown}
f(i) = \sin(i) \left[0.93\exp\left(\frac{-i^2}{2\sigma_1^2}\right) + 0.07\exp\left(\frac{-i^2}{2\sigma_2^2}\right)\right], \sigma_1=1.4^\circ, \sigma_2 = 17.0^\circ.
\end{equation}
As with the initial simulation in Section~\ref{s:sims}, the perihelion distances were evenly spread in the range $30 AU < q < 32 AU$.  This distribution of test particles is by no means meant to be a representation of the true CKB orbital distribution; instead it is merely a test of the applicability if our initial simulation results to a CKB-like orbital parameters and a realistic CKB inclination distribution.  

\begin{figure}
   \centering
   \includegraphics[width=6.75in]{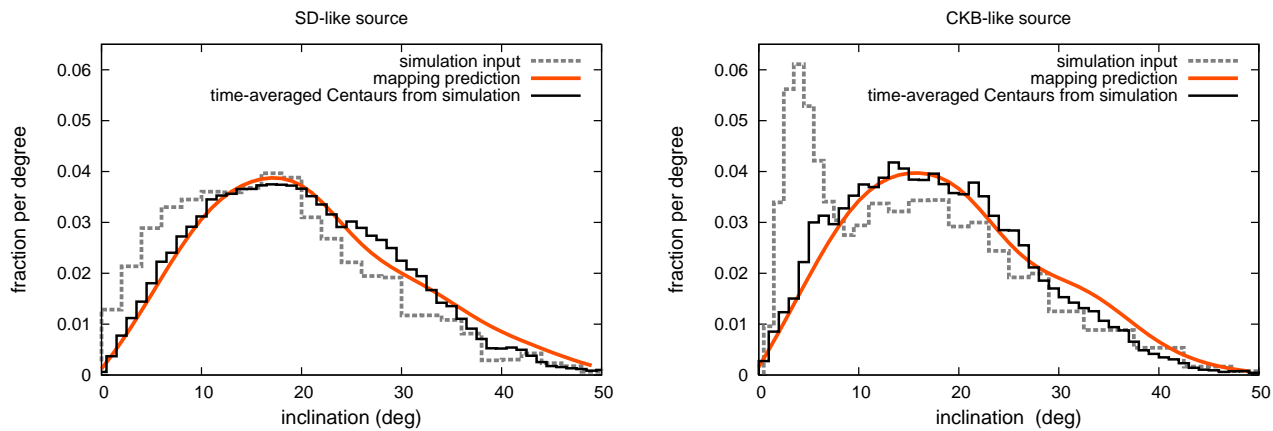} 
   \caption{Initial inclination distributions, the predicted Centaur distribution calculated from Eq.~\ref{eq:map}, and the time averaged Centaur distribution from a simulation for a SD-like source population with an inclination distribution given by Eq.~\ref{eq:sd} (left panel) and a CKB-like source population with an inclination distribution given by Eq.~\ref{eq:brown} (right panel).}
   \label{f:ckb_sd}
\end{figure}

The SD-like source region consists of 2400 test particles spread over a semimajor axis range $50 < a < 250$, perihelion range $30 AU < q < 32 AU$, and an inclination distribution given by:
\begin{equation}\label{eq:sd}
f(i) = \sin(i)\exp\left(\frac{-i^2}{2\sigma^2}\right), \sigma=17^\circ,
\end{equation}
which is the debiased best fit inclination distribution for the observed set of scattered disk objects \citep{Volk2008}.  Again, this SD-like source population is merely a test of the mapping procedure, not an accurate portrayal of the scattered disk distribution.  Both sets of source region test particles were integrated using the same simulation parameters described in Section~\ref{s:sims}.  Fig.~\ref{f:ckb_sd} shows the initial $i$-distributions for these simulations, the Centaur distribution predicted by the mapping procedure, and the time-averaged $i$-distribution for the Centaurs from the simulation.  The mapping procedure is quite successful in both cases, giving confidence that the mapping procedure can be used in lieu of detailed simulations to predict the Centaur $i$-distribution that would result from an arbitrary source region distribution.

\section{Results}\label{s:discussion}
\subsection{Inclination as a discriminator of Centaur sources}

It is apparent from the simulations that repeated encounters with the planets will significantly alter the inclination distribution of the Centaurs.  If there are any peaks in the source region's inclination distribution at $i<10^{\circ}$, these will generally not be preserved as the inclinations of the new Centaurs are raised by planetary encounters.  This is particularly relevant for the subset of Centaurs that might originate in the classical Kuiper belt, which has been shown to have a double peaked inclination distribution.  Even if Centaurs leak out of the CKB independently of their original inclinations, the double peaked nature of the inclination distribution will not persist in the Centaur region.  The reality is even less favorable for preserving any hints of the double peaked distribution, however.  In our model of the CKB from \citet{Volk2011}, we found that CKBOs from the high-$i$ portion of the inclination distribution were $\sim 4$ times more likely to evolve onto Neptune crossing orbits than CKBOs from the low-$i$ peak.  Consequently, the $i$-distribution of the Centaurs originating in the classical belt is nearly indistinguishable from the $i$-distribution of Centaurs originating in the scattered disk because the high-$i$ CKBOs and the SDOs have very similar inclination distributions;  in the case of our test simulations in Section~\ref{s:compare}, the Centaurs from the SD-like source region are nearly identical to those from the CKBO-like source region.

\begin{figure}
   \centering
   \includegraphics{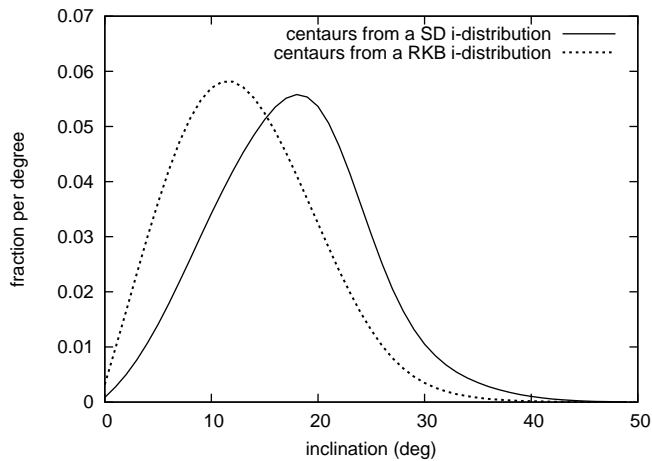} 
   \caption{Predicted Centaur inclination distributions calculated from Eq.~\ref{eq:map}, for the scattered disk source (solid line) and the resonant Kuiper belt source (dashed line).  The inclination distributions for the sources are taken from the debiased distributions published in \citet{Gulbis2010}.}
   \label{f:sd_res}
\end{figure}

The best chance to be able to identify source regions using the Centaur inclination distribution is if the proposed source regions have well separated high-inclination features, as these are better preserved than low-$i$ features.  Taking the debiased $i$-distributions from \citet{Gulbis2010} for several subclasses of TNOs as examples, the difference between the $i$-distribution of the scattered objects ($\mu = 19^{\circ}$, $\sigma = 7^{\circ}$) and the resonant objects  ($\mu = 6^{\circ}$, $\sigma = 9^{\circ}$) might be large enough to be observationally distinguishable in the Centaur supplied by these sources.  To examine the inclinations of Centaurs originating in the resonant Kuiper belt population, we integrated 600 test particles initially in the 3:2 mean motion resonance with Neptune with a uniform distribution of eccentricities in the range $0 < e < 0.3$ and inclinations $0 < i < 30^{\circ}$, consistent with the observed ranges of $e$ and $i$ for this resonance.  For the Centaurs generated in this simulation, we calculated their median dynamical lifetimes and average inclinations as a function of initial inclination in the resonance; the results are shown in Figs.~\ref{f:lt} and~\ref{f:icent}.  The simulated Centaurs from initially resonant orbits have lifetimes and inclinations that depend on initial inclination in the same way as the Centaurs originating from non-resonant orbits.  This allows us to apply the mapping procedure described in Section~\ref{s:map} to the inclination distribution of the resonant population (we use the debiased inclination distribution published by \citet{Gulbis2010}) to predict the inclination distribution of Centaurs originating from that population.   The result is shown in Fig.~\ref{f:sd_res}, in which we also plot the predicted distribution for Centaurs originating in the scattered disk population.   We note that the resonant source population yields a significantly lower peak inclination, $\sim11^\circ$, compared with that of the scattered disk source, $\sim18^\circ$.
Present observational data are insufficient to distinguish between these two distributions (cf.~Fig.~\ref{f:known_centaurs}). 
We estimate that the two predicted distributions in Fig.~\ref{f:sd_res} would be distinguishable given an observationally complete sample of approximately 100 Centaurs with $0 < i < 30^{\circ}$; this estimate is based on binning the distributions into $5^{\circ}$ bins, assuming Poisson noise statistics, and requiring that the $1\sigma$ error bars in the resulting histograms overlap in fewer than half the inclination bins.

\subsection{Retrograde and very high inclination Centaurs}
We mined our simulation results to identify cases where individual test particles evolved to very high inclinations in the Centaur region.  This is motivated by the discoveries of several Centaurs with very large inclinations, such as 2002 XU93 with $i \sim 78^{\circ}$ \citep{Elliot2005} and 2008 KV42,  with $i \sim 104^{\circ}$ \citep{Gladman2009}.  We examined our simulation results to see if test particles with lower $i_0$ (typical of TNOs) could be  excited to such large inclinations.  Our simulated Centaurs almost all started with $i_0 < 50^{\circ}$, and although some achieved inclinations as large as $80-90^{\circ}$ within the Centaur region, they did so only very briefly (timescales $<1$ Myr) before being ejected on hyperbolic orbits.  These also tended to have very large semimajor axes, $a>1000 AU$, when they were at large inclinations, which is inconsistent with the value $a\sim 40$ AU found for 2008 KV42.  We numerically integrated nearly 10,000 TNO-like test particles in the simulations presented in this work, and after following them all through many encounters with the planets, we find no Centaurs with orbits similar to the two previously mentioned high-$i$ Centaurs.  This result is consistent with the results of \citet{Brasser2012} who find the probability of evolving from a Kuiper belt-like orbit to a nearly retrograde Centaur orbit is $\sim10^{-5}$.  Therefore it appears unlikely that these relatively stable (stable on $\sim 10$ Myr timescales) observed high-$i$ Centaurs can be satisfactorily explained merely by typical orbital evolution of TNOs due to close encounters with the planets.  This can be understood quantitatively given that planetary encounters roughly preserve the Tisserand parameter, 
\begin{equation}
T = \frac{a_p}{a} + 2 \left[ \frac{a}{a_p}(1-e^2)\right]^{1/2}\cos i.
\end{equation}
Large changes in $\cos i$ (or a change in the sign of $\cos i$ as in the case of prograde to retrograde orbits) will tend to be associated with changes in $a$, possibly removing it from the Centaur region.  An alternate, very high-$i$ source region, as discussed by \citet{Gladman2009}, might be more likely.

\begin{figure}
   \centering
   \includegraphics[width=6.75in]{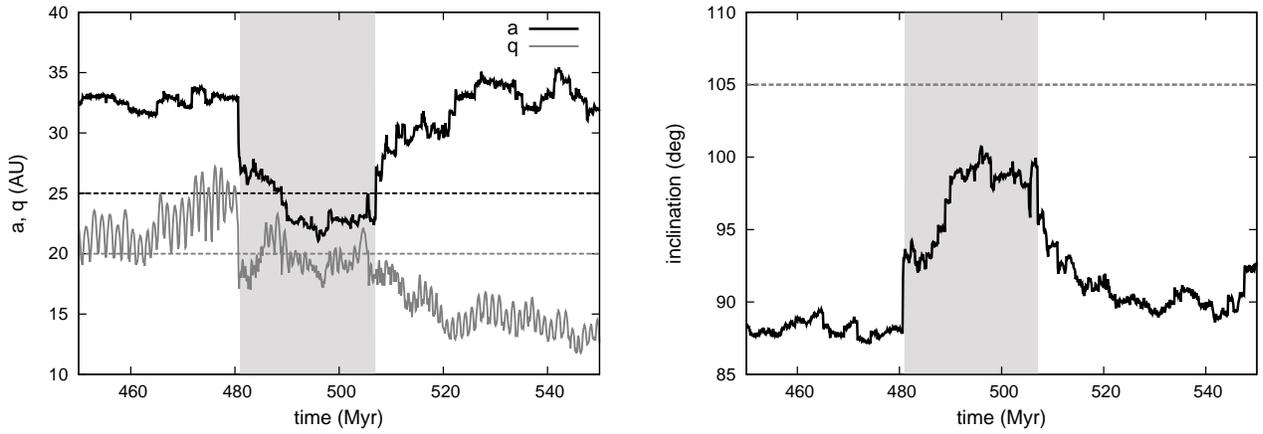} 
   \caption{Evolution of a test particle with $i_0=75^{\circ}$ onto a quasi-stable orbit (shaded region) similar to the observed retrograde Centaur 2008 KV42.  The values of $a$, $q$, and $i$ for 2008 KV42 are shown by the dashed lines while the evolution of the test particle is shown in solid lines.}
   \label{f:draclike}
\end{figure}

We performed an additional numerical simulation of test particles with very large initial inclinations,  $50 < i < 180^{\circ}$, semimajor axes in the range $50 < a < 250$ AU, and perihelion distances in the range $30 < q < 32$ AU; the simulation was run to $t=1$ Gyr.  We found that initially prograde test particles with $i_0 \gtrsim 70^{\circ}$ and $q>30$ AU were able to evolve into Centaurs ($5 < q < 30$ AU) with $i\sim100^{\circ}$.  As in our other simulations, most of these high values of $i$ were achieved only briefly while at very large values of $a$.  But we did find one case where a test particle with $i_0=75^{\circ}$ spent $\sim10$ Myr with an orbit very similar to that of 2008 KV42 ($a\sim25 AU, i\sim105^{\circ}, q\sim20 AU$); this case is shown in Fig.~\ref{f:draclike}.  This simulation demonstrates that there does exist a dynamical pathway from prograde, very high-$i$, scattered disk-like orbits to quasi-stable retrograde Centaur orbits.  Based on the $237$ initially prograde test particles that completed their evolution as Centaurs during this simulation, the probability of evolving from $50 < i < 90^\circ$ onto an orbit similar to 2008 KV42 is $\sim4\times10^{-3}$.  For the test particles with initially retrograde orbits ($90 < i < 180^{\circ}$), we find that the resulting Centaurs have average inclinations very similar to their initial inclinations; the median dynamical lifetime of our initially retrograde Centaurs is $\sim250$ Myr, compared to a median lifetime of $\sim145$ Myr for Centaurs originating from prograde, $i>50^{\circ}$ orbits.  A potential source for such very high-inclination and retrograde Centaurs is the Oort cloud, as recently suggested by \citet{Brasser2012}.  Our initial conditions for this simulation are consistent with the orbits of test particles that have entered the planet crossing region from the Oort cloud (see for example Fig. 2 in \citet{Brasser2012}).  The long dynamical lifetimes we find for these test particles in the Centaur region suggests that any contribution to the Centaur population from the Oort cloud (or any other source region with retrograde inclinations) should be long-lived compared to Centaurs originating from a prograde source region.

\section{Summary}
In summary, we have performed numerical simulations of the dynamical evolution of Centaurs originating from the Kuiper belt and the scattered disk and have found that:
\begin{itemize}
\item Multiple planetary encounters can significantly alter the inclination distribution of Centaurs relative to their source region's inclination distribution.\begin{itemize}
\item Test particles with $i_0$ $\lesssim$ $10^{\circ}$ tend to have their inclinations raised to values near $10^{\circ}$.
\item Test particles with $i_0$  $\gtrsim$ $10^{\circ}$ tend to have their inclinations lowered by a small amount that still preserves the correlation between average inclination and initial inclination.
\item  High-inclination features of the source region are better preserved than low-inclination features.

\end{itemize}
\item The median dynamical lifetimes of Centaurs increase nearly linearly with initial inclination.
\item The average absolute change in velocity of Centaurs due to planetary encounters can be represented as a power law, $|\Delta v| = 6.24\times10^3 (ms^{-1}) (\frac{M_p}{M_\sun})^{0.18} (\frac{a_p}{AU})^{-2.1} (\frac{r_{ce}}{R_H})^{-2}$.  
\item Planetary encounters are not an efficient way to produce extremely high-inclination Centaurs.  However, there is a dynamical pathway from the scattered disk at $i_0 \sim 70^{\circ}$ to retrograde Centaur orbits that remain retrograde for several tens of millions of years. 
\end{itemize}

\acknowledgments
\noindent This research was supported by grant no.~NNX08AQ65G from NASA's Outer Planets Research program.

\clearpage

\begin{deluxetable}{l l l l l}
\tabletypesize{\footnotesize}
\tablecolumns{4}
\tablewidth{0pt}
\tablecaption{Fit parameters for Eq.~\ref{eq:g} applied to the simulation results for each bin in initial inclination, $i_0$ (see Section~\ref{s:sims}), along with the median dynamical lifetime, $\tau$, of Centaurs from each bin in $i_0$.}
\tablehead{\colhead{$i_0$} & \colhead{$\mu$} & \colhead{$\sigma$} & \colhead{$\tau$ (Myr)}}
\startdata
$0-1^{\circ}$		& $0^{\circ}$		& $10.1^{\circ}$	& 23.8	\\
$1-2^{\circ}$		& $0^{\circ}$		& $9.1^{\circ}$		& 10.2	\\
$2-3^{\circ}$		& $0^{\circ}$		& $9.2^{\circ}$		& 14.1	\\
$3-4^{\circ}$		& $0^{\circ}$		& $10.8^{\circ}$	& 13.6	\\
$4-5^{\circ}$		& $0^{\circ}$		& $9.3^{\circ}$		& 12.0	\\
$5-6^{\circ}$		& $0^{\circ}$		& $10.0^{\circ}$	& 14.4	\\
$6-7^{\circ}$		& $0^{\circ}$		& $9.7^{\circ}$		& 17.1	\\
$7-8^{\circ}$		& $0^{\circ}$		& $9.5^{\circ}$		& 15.0	\\
$8-9^{\circ}$		& $0^{\circ}$		& $9.7^{\circ}$		& 23.6	\\
$9-10^{\circ}$		& $0^{\circ}$		& $10.2^{\circ}$	& 24.6	\\
$10-12^{\circ}$		& $0^{\circ}$		& $9.4^{\circ}$		& 22.5	\\
$12-14^{\circ}$		& $3.1^{\circ}$		& $9.3^{\circ}$		& 26.7	\\
$14-16^{\circ}$		& $7.4^{\circ}$		& $6.5^{\circ}$		& 33.7	\\
$16-18^{\circ}$		& $9.7^{\circ}$		& $6.6^{\circ}$		& 30.0	\\
$18-20^{\circ}$		& $12.0^{\circ}$	& $6.0^{\circ}$		& 46.6	\\
$20-22^{\circ}$		& $13.6^{\circ}$	& $6.4^{\circ}$		& 44.1	\\
$22-24^{\circ}$		& $16.6^{\circ}$	& $5.7^{\circ}$		& 41.6	\\
$24-26^{\circ}$		& $20.2^{\circ}$	& $4.0^{\circ}$		& 54.4	\\
$26-28^{\circ}$		& $20.3^{\circ}$	& $5.4^{\circ}$		& 53.9	\\
$28-30^{\circ}$		& $23.8^{\circ}$	& $4.6^{\circ}$		& 50.2	\\
$30-35^{\circ}$		& $27.5^{\circ}$	& $5.5^{\circ}$		& 57.7	\\
$35-40^{\circ}$		& $32.9^{\circ}$	& $4.9^{\circ}$		& 80.7	\\
$40-45^{\circ}$		& $38.0^{\circ}$	& $5.0^{\circ}$		& 101.4	\\
$45-50^{\circ}$		& $42.4^{\circ}$	& $5.1^{\circ}$		& 86.5	\\
\enddata
\label{t:fits}
\end{deluxetable}


\begin{thebibliography}{}

\bibitem[{{Bailey} \& {Malhotra}(2009)}]{Bailey2009}
{Bailey}, B.~L., \& {Malhotra}, R. 2009, \icarus, 203, 155

\bibitem[{{Barucci} {et~al.}(2011){Barucci}, {Alvarez-Candal}, {Merlin},
  {Belskaya}, {de Bergh}, {Perna}, {DeMeo}, \& {Fornasier}}]{Barucci2011}
{Barucci}, M.~A., {Alvarez-Candal}, A., {Merlin}, F., {Belskaya}, I.~N., {de
  Bergh}, C., {Perna}, D., {DeMeo}, F., \& {Fornasier}, S. 2011, \icarus, 214,
  297

\bibitem[{{Bernstein} {et~al.}(2004){Bernstein}, {Trilling}, {Allen}, {Brown},
  {Holman}, \& {Malhotra}}]{Bernstein2004}
{Bernstein}, G.~M., {Trilling}, D.~E., {Allen}, R.~L., {Brown}, M.~E.,
  {Holman}, M., \& {Malhotra}, R. 2004, \aj, 128, 1364

\bibitem[{{Brasser} {et~al.}(2012){Brasser}, {Schwamb}, {Lykawka}, \&
  {Gomes}}]{Brasser2012}
{Brasser}, R., {Schwamb}, M.~E., {Lykawka}, P.~S., \& {Gomes}, R.~S. 2012,
  \mnras, 420, 3396

\bibitem[{{Brown}(2001)}]{Brown2001}
{Brown}, M.~E. 2001, \aj, 121, 2804

\bibitem[{{Di Sisto} \& {Brunini}(2007)}]{DiSisto2007}
{Di Sisto}, R.~P., \& {Brunini}, A. 2007, \icarus, 190, 224

\bibitem[{{Di Sisto} {et~al.}(2010){Di Sisto}, {Brunini}, \& {de
  El{\'{\i}}a}}]{DiSisto2010}
{Di Sisto}, R.~P., {Brunini}, A., \& {de El{\'{\i}}a}, G.~C. 2010, \aap, 519,
  A112

\bibitem[{{Duncan} {et~al.}(1988){Duncan}, {Quinn}, \& {Tremaine}}]{Duncan1988}
{Duncan}, M., {Quinn}, T., \& {Tremaine}, S. 1988, \apjl, 328, L69

\bibitem[{{Duncan} \& {Levison}(1997)}]{Levison1997s}
{Duncan}, M.~J., \& {Levison}, H.~F. 1997, Science, 276, 1670

\bibitem[{{Elliot} {et~al.}(2005){Elliot}, {Kern}, {Clancy}, {Gulbis},
  {Millis}, {Buie}, {Wasserman}, {Chiang}, {Jordan}, {Trilling}, \&
  {Meech}}]{Elliot2005}
{Elliot}, J.~L., {et~al.} 2005, \aj, 129, 1117

\bibitem[{{Emel'yanenko} {et~al.}(2005){Emel'yanenko}, {Asher}, \&
  {Bailey}}]{Emelyanenko2005}
{Emel'yanenko}, V.~V., {Asher}, D.~J., \& {Bailey}, M.~E. 2005, \mnras, 361,
  1345

\bibitem[{{Everhart}(1972)}]{Everhart1972}
{Everhart}, E. 1972, \aplett, 10, 131

\bibitem[{{Fernandez}(1980)}]{Fernandez1980}
{Fernandez}, J.~A. 1980, \mnras, 192, 481

\bibitem[{{Fraser} \& {Brown}(2012)}]{Fraser2012}
{Fraser}, W.~C., \& {Brown}, M.~E. 2012, \apj, 749, 33

\bibitem[{{Fraser} {et~al.}(2010){Fraser}, {Brown}, \& {Schwamb}}]{Fraser2010}
{Fraser}, W.~C., {Brown}, M.~E., \& {Schwamb}, M.~E. 2010, \icarus, 210, 944

\bibitem[{{Gladman} {et~al.}(2009){Gladman}, {Kavelaars}, {Petit}, {Ashby},
  {Parker}, {Coffey}, {Jones}, {Rousselot}, \& {Mousis}}]{Gladman2009}
{Gladman}, B., {et~al.} 2009, \apjl, 697, L91

\bibitem[{{Gulbis} {et~al.}(2010){Gulbis}, {Elliot}, {Adams}, {Benecchi},
  {Buie}, {Trilling}, \& {Wasserman}}]{Gulbis2010}
{Gulbis}, A., {Elliot}, J., {Adams}, E., {Benecchi}, S., {Buie}, M.,
  {Trilling}, D., \& {Wasserman}, L. 2010, \aj, in press

\bibitem[{{Horner} \& {Lykawka}(2010)}]{Horner2010}
{Horner}, J., \& {Lykawka}, P.~S. 2010, \mnras, 402, 13

\bibitem[{{Horner} \& {Wyn Evans}(2006)}]{Horner2006}
{Horner}, J., \& {Wyn Evans}, N. 2006, \mnras, 367, L20

\bibitem[{{Levison} \& {Duncan}(1994)}]{Levison1994}
{Levison}, H.~F., \& {Duncan}, M.~J. 1994, \icarus, 108, 18

\bibitem[{{Levison} \& {Duncan}(1997)}]{Levison1997}
---. 1997, \icarus, 127, 13

\bibitem[{{Lowry} {et~al.}(2008){Lowry}, {Fitzsimmons}, {Lamy}, \&
  {Weissman}}]{Lowry2008}
{Lowry}, S., {Fitzsimmons}, A., {Lamy}, P., \& {Weissman}, P. 2008, in The
  Solar System Beyond Neptune, ed. M.~A. {Barucci}, H.~{Boehnhardt}, D.~P.
  {Cruikshank}, \& A.~{Morbidelli} (Tucson: Univ. of Arizona Press), 397--410

\bibitem[{{Morbidelli}(1997)}]{Morbidelli1997}
{Morbidelli}, A. 1997, \icarus, 127, 1

\bibitem[{{Peixinho} {et~al.}(2012){Peixinho}, {Delsanti}, {Guilbert-Lepoutre},
  {Gafeira}, \& {Lacerda}}]{Peixinho2012}
{Peixinho}, N., {Delsanti}, A., {Guilbert-Lepoutre}, A., {Gafeira}, R., \&
  {Lacerda}, P. 2012, \aap

\bibitem[{{Tegler} {et~al.}(2008){Tegler}, {Bauer}, {Romanishin}, \&
  {Peixinho}}]{Tegler2008}
{Tegler}, S.~C., {Bauer}, J.~M., {Romanishin}, W., \& {Peixinho}, N. 2008,
  {Colors of Centaurs}, ed. {Barucci, M.~A., Boehnhardt, H., Cruikshank, D.~P.,
  Morbidelli, A., \& Dotson, R.}, 105--114

\bibitem[{Tiscareno \& Malhotra(2003)}]{Tiscareno2003}
Tiscareno, M.~S., \& Malhotra, R. 2003, \aj, 126, 3122

\bibitem[{{Volk} \& {Malhotra}(2008)}]{Volk2008}
{Volk}, K., \& {Malhotra}, R. 2008, \apj, 687, 714

\bibitem[{{Volk} \& {Malhotra}(2011)}]{Volk2011}
---. 2011, \apj, 736, 11

\end{thebibliography}
\end{document}